# Allosteric Communication Pathways and Thermal Rectification in PDZ-2 Protein: A Computational Study.

Germán A. Miño-Galaz[*,1,2,3]

[1]Group of Nanomaterials. www.gnm.cl
Departamento de Física, Facultad de Ciencias, Universidad de Chile
Las Palmeras 3425, Ñuñoa, Santiago de Chile

[2]Centro Interdisciplinario de Neurociencias de Valparaíso (CINV),
Universidad de Valparaíso, Valparaíso, Chile

[3] Universidad Andres Bello Center for Bioinformatics and Integrative
Biology (CBIB), Facultad en Ciencias Biologicas, Santiago, Chile

[*]Corresponding Author Mailing address:
Republica 239, 3er piso, Santiago, Chile.

Phone: +562 2770 3612. E-mail: germino@u.uchile.cl;
german.mino.galaz@gmail.com. Website: www.gnm.cl.

## Abstract

Allosteric communication in proteins is a central and yet unsolved problem of structural biochemistry. Previous findings, from computational biology (Ota and Agard, 2005), have proposed that heat diffuses in a protein through cognate protein allosteric pathways. This work studied heat diffusion in the well-known PDZ-2 protein, and confirmed that this protein has two cognate allosteric pathways and that heat flows preferentially through these. Also, a new property was also observed for protein structures - heat diffuses asymmetrically through the structures. The underling structure of this asymmetrical heat flow was a normal length hydrogen bond (~2.85 Å) that acted as a thermal rectifier. In contrast, thermal rectification was compromised in short hydrogen bonds (~2.60 Å), giving rise to symmetrical thermal diffusion. Asymmetrical heat diffusion was due, on a higher scale, to the local, structural organization of residues that, in turn, was also mediated by hydrogen bonds. This asymmetrical/symmetrical energy flow may be relevant for allosteric signal communication directionality in proteins and for the control of heat flow in materials science.

## Introduction

Allosteric communication in proteins is a central and yet unsolved problem in structural biochemistry. This phenomenon is involved in crucial cellular and physiological functions and is determinant in human diseases.[1-3] Allosteric communication can mediate signal transmission through proteins on both short (3 Å) and long (100 Å) range scales, and is studied using both experimental[4-12] and theoretical [8,13-19] techniques, and its occurrence has been clearly established.

Several questions surround the phenomenology of signal transduction and energy flow in proteins. This paper assesses how energy is transported from one site to another in proteins. In particular, directionality for energy flow in proteins was empirically observed. The structure underlying this directional energy flow was the normal length hydrogen bond (~2.85 Å). This result suggests that a normal length hydrogen bond is the minimal chemical structure that can operate as a thermal rectifier in biomolecules. Another source of asymmetrical heat diffusion was found on a higher scale in the local structural organization of residues, which were also mediated by hydrogen bonds. In contrast, the thermal rectification effect seemed to be suppressed in short hydrogen bonds (~2.60 Å), giving rise to symmetrical thermal diffusion. Asymmetrical energy flow may be relevant for allosteric signal communication directionality in protein structures and for heat flow control in the field of materials science.

Hydrogen bonds have been an inexhaustible source of research for decades.[20] Their role as stabilizing agents in protein structures has been clearly established.[21] They also act in the modulation of chemical reactivity in proteins[22-24] and, as has been proposed earlier in the field of physics, can act as a supporting structure for vibrational energy flow in protein structures.[25] Recent studies have shown that hydrogen bonds are also important in heat diffusion across the β-sheet structure of the spider silk protein[26] and in α-helices.[27]

Due to their polymeric nature, when proteins fold to reach their biologically active state they generate a complex network of contacts. Previous reports have shown that thermal energy flows in proteins according to the physical connectivity of this network[28], with velocities of propagation on the order of 10 Å ps$^{-1}$.[14] This flow of thermal energy mimics the heat transport of percolation clusters, where the energy flows anisotropically, that is, faster along physically connected channels and slower along the numerous pathways that reach dead-ends.[14] The anisotropic heat flow in proteins was empirically demonstrated by Ota and Agard using a computational pump-probe method termed Anisotropic Thermal Diffusion (ATD).[30] Anisotropic thermal diffusion is a non-equilibrium molecular dynamics (MD) pump-probe method that has been used to study allosteric communication pathways in proteins.[15,29-30] Various studies support a direct correlation between the heat diffusion pathways in proteins and their known allosteric communication pathways.[15,29-30] In its original formulation,[30] the protein model is equilibrated in the absence of a solvent at a very low temperature to minimize background noise. After this, a specific residue is heated by coupling it to a thermal bath (pump), and the increase in vibrational energy is measured in different sites of the protein (probe) over a short time lapse. The correlations between anisotropic vibrational energy propagation and the cognate structural allosteric pathways have been confirmed in other systems,[31] supporting the utility of ATD for the characterization and comprehension of these mechanisms.

In the simplest cases, allosteric communication involves a protein, a ligand site, and an effector site in the same protein. After ligand binding, a signal is propagated throughout the protein structure to the effector site. Depending on the response of the signal on the effector site, allosteric modulation may be positive or negative.[32] As an example of positive modulation, the binding of vancomicin to the glycopeptide antibiotic facilitates the dimerization of two molecules of the antibiotic.[33] On the other hand, negative modulation is observed when cAMP molecules bind to the catabolite activator protein (CAP). Here, the binding of the first cAMP molecule makes the binding of the second cAMP to CAP more difficult.[10]

Thermodynamically, the free energy of allosteric processes in proteins has been interpreted by attributing the enthalpic part to conformational changes in the protein structure and the entropic contribution to the dynamical fluctuations of residues.[7,34-35] Based on this distinction, three types of allosteric signaling are distinguished: Type I, with small or subtle conformational changes, governed largely by entropy ($\Delta H \approx 0$); Type II, with the participation of enthalpic and entropic components to different extents; and Type III, largely governed by enthalpic changes ($\Delta S \approx 0$).[34] Type I allosteric communication has been less studied as compared to the enthalpic communication of Types II and III.[5-6] Type I allosterism, also known as dynamic or configurational allosterism, was initially proposed by Cooper and Dryden[35] in 1984. According to these authors, Type I allosterism arises from changes in the frequencies and amplitudes of macromolecular fluctuations around the same protein conformation and involves several forms of dynamic behavior, ranging from highly correlated low-frequency normal-mode vibrations to random local anharmonic motions of individual groups or atoms. These fluctuations can be described in terms of vibration, libration, or rotation of individual groups that propagate vibrational energy along the protein structure.[35] Systems such as staphylococcal nuclease,[36] myoglobin,[37] serine proteases,[38] HIV-1 protease,[39] dihydrofolate reductase,[40] β–lactamase,[41] and the Post–synaptic density-95/Drosophila disc/Zonula occludens-1 (PDZ) domain protein family[7,16,30] are known to transfer allosteric information by altering their dynamics without significant conformational changes (Type I allosterism).

The allosteric pathways of the PDZ domain family are also interesting as they involve residues which are distant in terms of protein structure. The PDZ protein family has a common structural domain of 80-96 amino-acids, and appears in signaling proteins of bacteria, yeast, plants, viruses,[42] and animals.[43] Experimental,[7] statistical coupling analysis (SCA),[16] and theoretical models[44] have been used to characterize this protein family, and a consensus of its allosteric pathways has been reached. In particular, using cross-correlation analysis, Kong and Karplus[19] report two pathways of allosteric communication for

PDZ-2. In Pathway I (Figure 1), the allosteric communication pathway goes from strand $\beta_2$ (residues 19-24) and extends along the long axis of the helix $\alpha_1$ (residues 44-49), and Pathway II starts at strand $\beta_2$ and goes across strands $\beta_3$ (residues 33-40), $\beta_4$ (residues 56-61), and $\beta_6$ (residues 83-90) to $\beta_1$(residues 6-13). A structural interpretation for the correlations obtained by SCA[16] was initially obtained using ATD.[30] The pathways of energetic connectivity observed with ATD were consistent with the correlations between residues found by SCA.[16,30]

As ATD is a quick method, it can be extended to the automatic mapping of heat propagation from every residue in a protein structure.[31] In this numerical study, a review of the heat propagation pathways of the PDZ-2 protein was comprehensively performed, and full heat propagation maps of energetic connectivity were obtained. Confirmation that vibrational energy flows through the connectivity network via previously reported[19] allosteric pathways is offered, but it is also shown that these paths of energetic connectivity were not found to be completely symmetric. The observations suggest that normal length hydrogen bonds can act as thermal rectifiers for heat diffusion and that this rectifying capability can be modulated by bond length. The study and understanding of heat flow directionality – phonon rectification – is highly desirable in the field of materials science for the development of thermal gates,[45] thermal memories,[46] and acoustic cloakers.[47] In this respect, the structural inclusion of the rich dynamics of hydrogen bonds in the designs of materials science may be a useful alternative for implementing these thermal devices.

**Methods and Computational Details.**

The PDZ-2 protein in the presence of the ligand,[19] herein termed PDZ-2·L (PDB ID: 1D5G), and without the ligand, herein termed PDZ-2 (PDB ID: 3PDZ), were used for simulations. These models were minimized and equilibrated using periodic boundary conditions at 298 K and 1 atm with water as an explicit solvent and NaCl at 0.15 M. Constant temperature conditions were used through

Langevin dynamics with a damping coefficient of 5 ps$^{-1}$. An integration time-step of 2 fs and a cutoff of non-bonded forces at 12 Å with a switching function starting at 10 Å were used. Equilibration was divided into three steps. First, the coordinates of the protein models were fixed to allow the solvent and ions to equilibrate for 200 ps. Second, the coordinates of the lateral chains were released and equilibrated for another 200 ps. Third, all atoms were released and allowed to equilibrate for 2 ns. Finally, a production run of 20 ns was performed, after which 20 equally spaced frames were selected for the ATD procedure. For each of the 20 frames, the solvent was removed and the protein model was cooled to 10 K for 100 ps MD. At this point, the ATD procedure was applied using a set of automated scripts[31] available at https://leandro.iqm.unicamp.br/atd-scripts and by adding the option rigidbonds=none. The scripts allowed the automated, individual heating of each residue of the protein, in addition to permitting temperature analysis of every other residue after a short predefined time-lapse of 30 ps. Each residue was individually coupled to a thermal bath at 300 K, and the thermal response in the rest of the protein was measured. A total of 20 independent ATD scans were performed for PDZ-2 and PDZ-2·L. The results were averaged and used to construct the contact and thermal diffusion maps shown in Figure 2. All simulations were performed using the NAMD2.9 software[48] with the CHARMM22[49] potential and CMAP correction[50]. Visualization and data analysis were performed using the VMD 1.8.7 software.[51]

**Results.**

Figures 1 (a) and (b) show the structures of PDZ-2 and PDZ-2·L. Figures 2 (a) and (b) show the contact and thermal diffusion maps averaged for 20 independent simulations for both cases. The thermal diffusion maps showed a rough similarity with the connectivity maps. These plots showed that heat diffused mainly through structure connectivity. These contacts belonged to the covalent network of the primary structure or arose from the side-chain interactions of residues in different elements of the secondary structure of the models. The thermal maps revealed that the allocation of allosteric Pathways I and II for both models can be recovered using ATD, and both pathways can be deduced by inspecting the averaged thermal maps. A close inspection of Figure 2 uncovered how Pathway II was recovered. By looking at the abscissa of the thermal map for PDZ-2 at $\beta_2$, a thermal coupling to $\beta_3$ was observed (Box 1). Next, looking at the thermal map for the abscissa at $\beta_3$, a vibrational coupling with $\beta_4$ was seen (Box 2). Then, entering the thermal map for the abscissa at $\beta_4$, a vibrational coupling with $\beta_6$ appeared (Box 3). Finally, by entering the abscissa at $\beta_6$, a strong vibrational coupling to $\beta_1$ was noted (Box 4). Now, in order to recover Pathway I, the PDZ-2·L thermal map was analyzed. By entering in the abscissa at $\beta_2$, a vibrational coupling with $\alpha_1$ occurred (encircled area). In the case of PDZ-2 a strong coupling among $\beta_2$ and $\alpha_1$ was not found. Thus, the presence of the ligand enhanced this pathway. This can be clearly observed by comparing both thermal maps. With this inspection, Pathways I and II[19] were reproduced.

Vibrational couplings in other regions were affected by the presence of the ligand. The thermal map of PDZ-2 at abscissa in $\beta_3$ displayed a strong vibrational coupling to residues 66-69 (a region between $\beta_5$ and $\alpha_2$) and to residue 90 of $\beta_6$. Both these thermal coupling were quenched in the presence of the ligand. Therefore, heat propagation was once again found modulated by the presence of the ligand. This modulation resulted from observable structural rearrangements, but these were not evident from the contact maps alone. The similarity between contact maps and thermal maps suggests that communication pathways in Type I allosteric systems are primarily defined by atomic connectivity and its respective interaction. In this particular system (PDZ-2), connectivity occurred among secondary structure elements of the protein.

The structural features that gave rise to the selected energetic couplings were analyzed in detail. For example, the thermal maps in Figure 2 displayed a thermal coupling between $\beta_2$ and His71, corresponding to $\alpha_2$ (highlighted with a triangle in the thermal maps). In the ligand-free PDZ-2 model (Figure 3a), this coupling arose from alternating hydrogen bonds between N$\delta$1 of His71 with O$\gamma$1 or with the oxygen atom of the C=O backbone moiety of Thr23 (Figure 3c, top).

This coupling also occurred from hydrogen bonds between Nδ1 of His71 with Oγ1 or with the oxygen atom of the C=O backbone moiety of Thr23 (Figure 3c, bottom). This effect was also observed in the thermal map of PDZ-2·L (Figure 3b). In this case, the heat transference between $\beta_2$ and $\alpha_2$ was also due to the formation of hydrogen bonds, but it occurred between Nδ1 of His71 and the oxygen atom of the C=O backbone moiety of Gly24 (Figure 3d, top) or between Nε2 of His71 and Oγ1 of Thr23 (Figure 3d, bottom). These observations were consistent with recent reports about the role of hydrogen bonds in heat diffusion in protein structures.[26,27]

Further observation of the results presented in Figure 2 revealed interesting features of heat propagation. Asymmetries were observed in the thermal diffusion maps. This was qualitatively observed using the color scale of the thermal maps in Figure 2. For instance, in Pathway II the heat interchange between $\beta_6$ and $\beta_1$ was asymmetrical (observed in both PDZ-2 and PDZ-2·L). Likewise, the heat interchange in Pathway I between $\beta_2$ and $\alpha_1$ was also asymmetrical (observed in PDZ-2·L). It is quite possible that there is a preferential directionality in both cases. These asymmetries were observed at other points on the averaged thermal maps and supported the idea of signal propagation directionality in allosteric pathways. These asymmetries were rationalized by analyzing structure, pump-probe plots, and energy flow, as shown in the following cases.

Now, the asymmetrical interaction between Asp5 and Lys91, which are located in $\beta_6$ and $\beta_1$ respectively, of the PDZ-2 case will be presented. These two residues interacted through a backbone hydrogen bond (Figure 4a). A clear asymmetrical pump-probe relationship was observed in Figure 4b (circles), in which heat was transferred with a preferential directionality going from Lys91 towards Asp5. In the 20 propagations (see Methods section), 63% of time these residues interacted through one backbone hydrogen bond; 4.5% of the time residues interacted between hydrogen bonds formed by the lateral chains, and 36.5% of the time no interaction was observed (distances greater than 3.5 Å). The pump-probe profile with standard deviation (Figure 4c) revealed a response at Asp5 of 98.4 ± 15.9 K when the pump was applied to Lys91, and a response at

Lys91 of 72.5 ± 12.3 when the pump was applied to Asp5. These data, in addition to the structural inspection of the hydrogen bond interaction, suggest that the preferential directionality of the heat flow (Lys91 → Asp5) is due to the spatial orientation of the hydrogen bond, that, in this case, has the configuration Lys91-N-H···O=C-Asp5 (Figure 4a, right). To evaluate the possible relationship between heat flow directionality and orientation of the hydrogen bond, the kinetic energy in time of the atoms involved was determined and shown for three heat diffusion trajectories (Figures 4d, 4e, and 4f). As can be observed in Figure 4d, when the pump was applied to Lys91 a kinetic energy of ~1000 J/mol was observed at 10000 fs for the oxygen atom of the C=O moiety of Asp5 (Figure 4d, left). Inversely, when the pump was applied to Asp5 a kinetic energy of ~750 J/mol at 10000 fs was observed for the nitrogen atom of the N-H moiety of Lys91 (Figure 4d, right). The same observations held in Figures 4e and 4f, in which the C=O moiety consistently reported increased kinetic energy along the propagations. These results imply that in this particular configuration of the hydrogen bond, heat diffuses easily in the direction N-H → O=C, and with difficultly in the inverse direction, N-H ← C=O. The origin of this effect may be due to the structural asymmetry of the hydrogen-bond and to the consequent stretching force constants associated with the N-H and C=O bonds (Figure 4a, left). For N-H and C=O, the stretching force constants were 440 and 620 kcal·mol$^{-1}$·Å$^{-2}$, respectively. Each moiety also had a particular partial charge state that was linked to the force field and the associated vibrations, which may give rise to the observed phenomenology. Further analysis is required to give an exact explanation of the effect, which is beyond the scope of this article. Based on the observations for this specific case, it appears that the normal hydrogen bond acts as a thermal rectifier, that is, it can act as a molecular motif that transfers thermal energy with preferential directionality. This phenomenon is not new and has been observed in hydrogen-bond enhanced thermal energy transport in functionalized, hydrophobic, and hydrophilic silica−water interfaces.[52] Nevertheless, the present report is, to the author's knowledge, the first to demonstrate this phenomenon in protein structures. In the case of PDZ-2·L, a similar phenomenon was observed when analyzing the interaction between Asp5-Lys9, but due to the higher percentage of interaction through lateral chains for the PDZ-2·L, subsequent discussion is, at this point, restricted only to PDZ-2.

In the case of symmetrical heat diffusion, attention was directed to the structural arrangement of Glu10-His86-Glu8 in PDZ-2. This arrangement interacted through two hydrogen bonds (Figure 5a). In this specific case, hydrogen bonds were intermittently formed between the available proton donors located at His86 (atoms N$\delta$1 and N$\varepsilon$2) and the respective proton acceptors (Atoms O$\varepsilon$1 and O$\varepsilon$2) located at carboxilyc moieties of the Glu10 and Glu8 residues. In the 20 propagations, >95% presented these interactions. A symmetrical pump-probe relation was observed in the interaction of Glu10-His86 and His86-Glu8 for PDZ-2 (Figure 5b, rectangle). The pump-probe plots showed responses at Glu8 and Glu10 of 114.3 ± 11.7 and 120.6 ± 8.6 K, respectively, when His86 was excited (Figure 5c, left). Inversely, a response at His86 of 120.0 ± 12.6 K was observed when Glu10 was excited (Figure 5c, right). Finally, a response at His86 of 115.4 ± 7.2 K was observed when Glu8 was excited (plot not shown). These measurements reinforce the idea of an asymmetrical pump-probe relationship. Evaluations of the kinetic energy of the hydrogen bonds for the interaction between Glu10 and HIS86 in three selected trajectories are shown Figures 5d, 5e, and 5f. On one hand, when the pump was applied to Glu10 a kinetic energy of ~1500 J/mol at 10000 fs was observed for the nitrogen atom of the N-H moiety of His86 (Figures 5d, 5e, and 5f, left). These measurements showed a higher value when compared to the kinetic energy analysis when Asp 5 was excited, in which a response of ~750 J/mol at 10000 fs for the nitrogen atom of the N-H moiety of Lys91 was found (Figures 4d, 4e, and 4f, left). On the other hand, when His86 was excited, similar kinetic patterns were observed with respect to the heat injection to Glu10. Comparison of the kinetic energy plots with the previous Lys91-Asp5 case supports the idea of a symmetric response of the Glu10-His86 arrangement. The main difference in both cases is that the former formed a charge-assisted hydrogen bond with an average length of ~2.62 Å (Figure 5, see insets in kinetic energy analysis), while the latter formed a normal non-charged hydrogen bond with an average length of ~2.82 Å (Figure 4, see insets in kinetic energy analysis). An enhanced chemical reactivity of structural elements that surround hydrogen bonds that can suffer transition from normal bonding (> 2.8 Å) to a short bonding regime (< 2.6 Å) has been previously reported.[23] In line with this, the results of this report suggest that heat diffusivity behavior, whether symmetric or asymmetric, may be modulated by the length of the hydrogen bond.

The asymmetrical pump-probe relationship between the residues His32 and Glu90 was also analyzed for the ligand-free PDZ-2 model. These residues belong to Pathway II. As shown in Figure 6a, these residues were located in the loop connecting $\beta_2$-$\beta_3$ and at the end of the $\beta_6$ strand. A detailed representative snapshot of the residues that were in contact with His32 and Glu90 is depicted in Figure 6b. As can be observed in the detail of the thermal map (Figure 6c), a preferential directionality of heat flow existed, in which the heating of Glu90 had an increased pump to His32. The pump-probe plots of His32 and Glu90 (Figures 6c and 6d) revealed an effective asymmetric relationship between these two residues. While the heating of Glu90 raised the local temperature of His32 up to 92.7±17.2 K, the heating of His32 raised the local temperature of Glu90 to 74.5±17.7 K. Analysis of Figure 6b directly explained the results of Figure 6d. When the pump was applied at Glu90, heat mainly diffused to Arg57, His32, and Asp5. On the other hand, when the pump was applied at His32, heat diffused to Arg57 and Glu90. Given that His32 had contact with Glu67 at two alternating or simultaneous points (Figure 6b), increased quantities of vibrational energy were transferred to His32, thus quenching the transference to Glu90. Here Glu67 was the most important heat sink, giving rise to the asymmetrical behavior. Charge-assisted hydrogen bonds were formed between Glu90, Arg57, and His32, while His32 formed a charge-assisted hydrogen bond with Glu67. It seems that the counter balance of the contacts gave rise to the observed behavior in this case.

Another example of heat transfer directionality was observed in Pathway I. This pathway was observed in PDZ-2 and was enhanced in the case of PDZ-2·L (Figure 2, encircled area). This involved the group of residues Lys13, Asn14,

Asp15, Asn16, and Ser17 located in the loop connecting $\beta_2$-$\beta_6$, as well as Gln43 located in the loop connecting $\beta_3$-$\alpha_1$ (Figure 7a). A particular feature of group 13-17 was the formation of a hydrogen bond cluster which formed a loop in this region (Figure 7b). It seems that this cluster readily exchanged heat within itsel, through the hydrogen bond cluster, and it then transferred thermal energy to Gln43. On the other hand, Gln43 was only able to transfer back to Asp15, as can be observed in the thermal map detail in Figure 7c. In this case, the source of asymmetric heat diffusion was given by the local three-dimensional organization of the residues that formed a funnel-like structure pointing towards Gln43, thus facilitating heat transfer to it, naturally impeding the inverse process. For completeness, the statistical pump-probe behavior is shown in Figure 7d.

In summary, the results showed that it is possible to recover Pathways I and II using ATD. These observations suggest an important role of normal length hydrogen bonds in the directionality of vibrational energy transfer in proteins. Moreover, the topological organization of residue contacts is proposed as an important source of directional energy transfer in biomolecules.

## Discussion and Conclusions.

Signal transduction in proteins is an important problem in structural biochemistry.[1-19] The aim of this work was to test one methodology, ATD[30], in order to identify pathways of energy propagation in the PDZ domain and to correlate them with the cognate I and II allosteric pathways for PDZ-2. This protein transfers allosteric signals with minor conformational changes[7,9,12,19,53] and has been cataloged as a Type I entropic allosteric system.[34] This type of allosteric communication is manifested through alterations in the frequencies and amplitudes of residue fluctuations and involves several forms of dynamic behavior, from highly correlated, low-frequency normal-mode vibrations to random local anharmonic motions of individual groups or atoms.[33,35] Evidence from experimental studies,[7] SCA,[16] and theoretical analysis[19,44] supports the existence of two allosteric pathways for PDZ-2. In Pathway I (Figure 1), allosteric communication went from strand $\beta_2$ and extended along the long axis of the helix $\alpha_1$. Pathway II started at strand $\beta_2$ and went across strands $\beta_3$, $\beta_4$, and $\beta_6$ to $\beta_1$.

The results demonstrate that PDZ-2 and PDZ-2·L have similar contact maps that resemble the thermal diffusion maps. This study concludes that a correlation exists between the flow of vibrational energy, the cognate I and II allosteric pathways for PDZ-2 and PDZ-2·L, and their respective contact maps. Thus, the connectivity between secondary elements of the protein structure appears to be the determining factor in Type I allosteric communication. The presence of a ligand can induce changes in the contacts between secondary elements so that the ligand can act as a thermal switch for the vibrational coupling of the allosteric pathways.

The results also show that hydrogen bonds are implicated in the heat flow among different segments of the protein, as was found in the coupling between $\beta_2$ and $\alpha_2$. In this case, hydrogen bonds were formed between His71 and Thr23 for PDZ-2, and between His71 and Thr26 or between His71 and the C=O backbone moiety of Gly24 for PDZ-2·L. The role of hydrogen bonds in enhancing heat diffusion has been demonstrated in models of $\alpha$-helices,[27] in which a higher number of hydrogen bonds produces higher thermalization rates. The role of hydrogen bonds in heat diffusion has also been shown in spider silk.[26] Thus, the observations of the present study are consistent with and supported by these previous reports.

The main result of this research was the detection of preferential heat flow directionality along allosteric pathways. This directionality was observed through the details provided by the thermal maps for PDZ-2 and PDZ-2·L. Directionality occurred in the absence of a ligand and could be modulated by its presence. This phenomenon was studied at points of PDZ-2 and was consistent with the expected behavior of an allosteric protein, that is, to transfer signals with preferential directionality. The smallest organizational unit that presents this property is the hydrogen bond. A punctual analysis of the interaction between Asp5 and Lys91 suggested that normal length hydrogen bonds behave as thermal rectifiers, which is to say, they transfers vibrational energy with preferential directionality. This effect may originate from the natural asymmetry of the hydrogen bond given by the N-H and C=O moieties. Also, because N-H is a single bond and C=O is a double bond, the amplitude of the stretching that each moiety reaches during similar thermal excitation is higher for the latter and lower for the former. Further research is required to reveal the mechanism of the asymmetrical heat diffusion in this particular structure. The role of hydrogen bonds as thermal diodes has been previously proposed in MD simulations of asymmetrical functionalized silica slabs floating in water.[52] In these simulations, one side of the silica is hydrophobic and the other hydrophilic. Heat flow is favored when applied in the hydrophobic to hydrophilic direction and impeded in the inverse direction. On a smaller scale, this research shows that the heat flow is favored in the direction of N-H → O=C and impeded in the inverse direction. This behavior is compromised in the case of short hydrogen bonds, as was observed in the pump-probe analysis of the interaction of Glu10-His86. As Glu and His have -1 and +1 charges, respectively, both assisted the formation of short hydrogen bonds, ~2.6 Å. This was the main difference in regards to the interaction between Asp5-Lys91, which presented a normal hydrogen bond of length of ~2.85 Å. It seems that the short distances between the N-H and the C=O moieties observed for the interaction between Glu10-His86 impeded the asymmetric behavior of thermal diffusion. Quantum mechanical models have previously shown that the transition between a normal and short bonding regime significantly alters the chemical reactivity of the structural elements that surround a given hydrogen bond.[23] The present study, using classical MD, suggests that the transition from a normal to a short hydrogen bond regime may modulate the thermal diode effect.

Asymmetrical heat transference was also observed in the complex interaction of residues, as depicted in Figures 6 and 7. In the case of the asymmetrical pump-probe relationship between Glu90-His32, a set of hydrogen-bonded interacting residues explains the observed directionality. For example, when heat was provided to Glu90, it was directly transferred to His32, Arg57, and to Asp5, whereas the structural organization surrounding His32, namely, Gly33 and Glu67, quenched the heat transfer to Glu90. As shown in the pump-probe plot of His32, Glu67 acted as a thermal sink that trapped an important quantity of heat. So, the topological organization of a set of interacting residues may be the reason for asymmetrical heat diffusion behavior. A similar phenomenology seems to be operative in the interaction of residues 13-17 and Gln43 (Figure 7). Observation of the structure at this segment revealed a hydrogen bond network organization similar to a funnel pointing towards residue Gln43. As residues 13-17 were heated, they distributed heat among each other and transferred it to Gln43, while the heating of Gln43 mainly transferred back to residue Asp15. This residue was the individual heat connection of group 13-17 to Gln43. All the previous data suggest an important role of hydrogen bonds in vibrational coupling among segments of the secondary structure in this system and, therefore, in the constitution of allosteric pathways.

In summary, this work analyzed the vibrational energy diffusion in PDZ-2 and the correlation with cognate allosteric pathways for this system. Examples of asymmetrical heat propagation in protein structures that give rise to signal directionality were also offered. Thermal diffusion maps showed a high degree of similarity with connectivity maps, thus supporting the idea that Type I allosteric communication is the connectivity network that defines the allosteric route and its directionality. Taken together, it is suggested that hydrogen bonds play an important role as a transport structure for heat and thermal rectification in the allosteric communication pathways of proteins.

## References.

## Acknowledgments.

The author would like to acknowledge funding provided by Fondecyt Grant 3110149 awarded to him and by partial support from grants (i) Millennium Initiative P09-022-F, (ii) Conicyt Proyecto Anillo ACT – 1104, and (iii) Fondecyt 1131003.

The author would also like warmly thank Professor Leandro Martinez for his support and advice during research. Grants FAPESP (Brazil) Projects 2010/16947-9 and 2013/05475-7.

**Figures and Captions.**

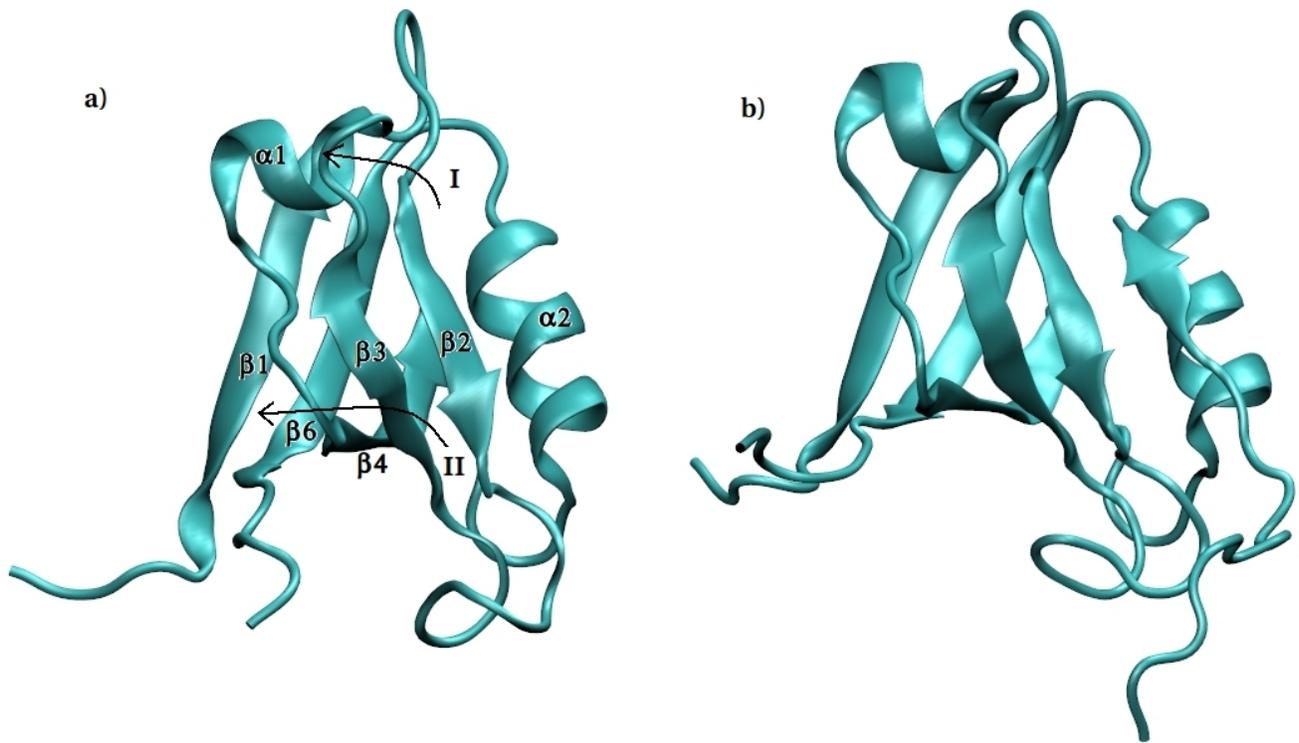

**Figure 1.** PDZ-2 with (a) and without ligand (b). In (a) the proposed Pathways I and II are depicted. Beta sheet 5 is located behind β2 in both images.

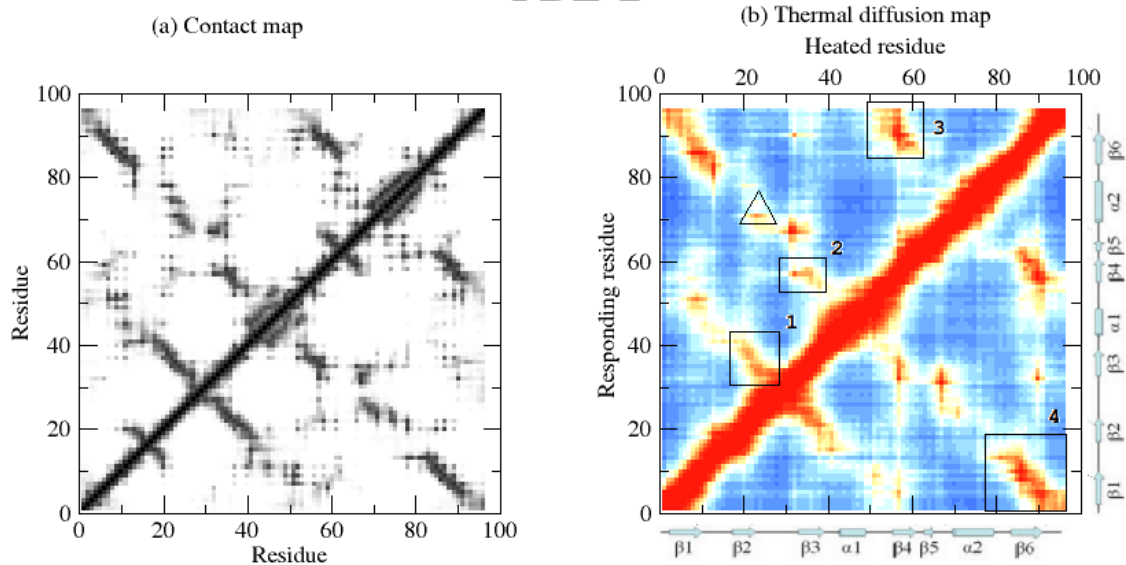
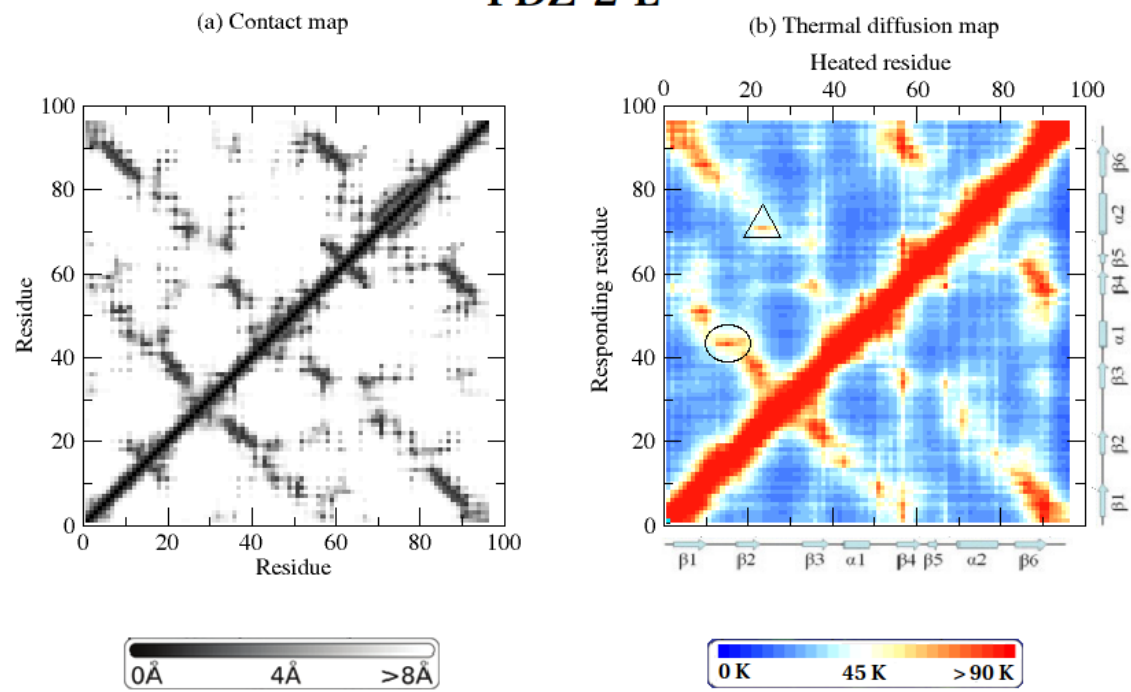

**Figure 2.** Average contact (a) and thermal diffusion map (b) for PDZ-2 and PDZ-2·L (n=20). Boxes 1, 2, 3, and 4 represent coupling associated with Pathway II. Circles represents coupling associated with the pathway and triangles represent a case of heat transference mediated by hydrogen bonds (see text for further details).

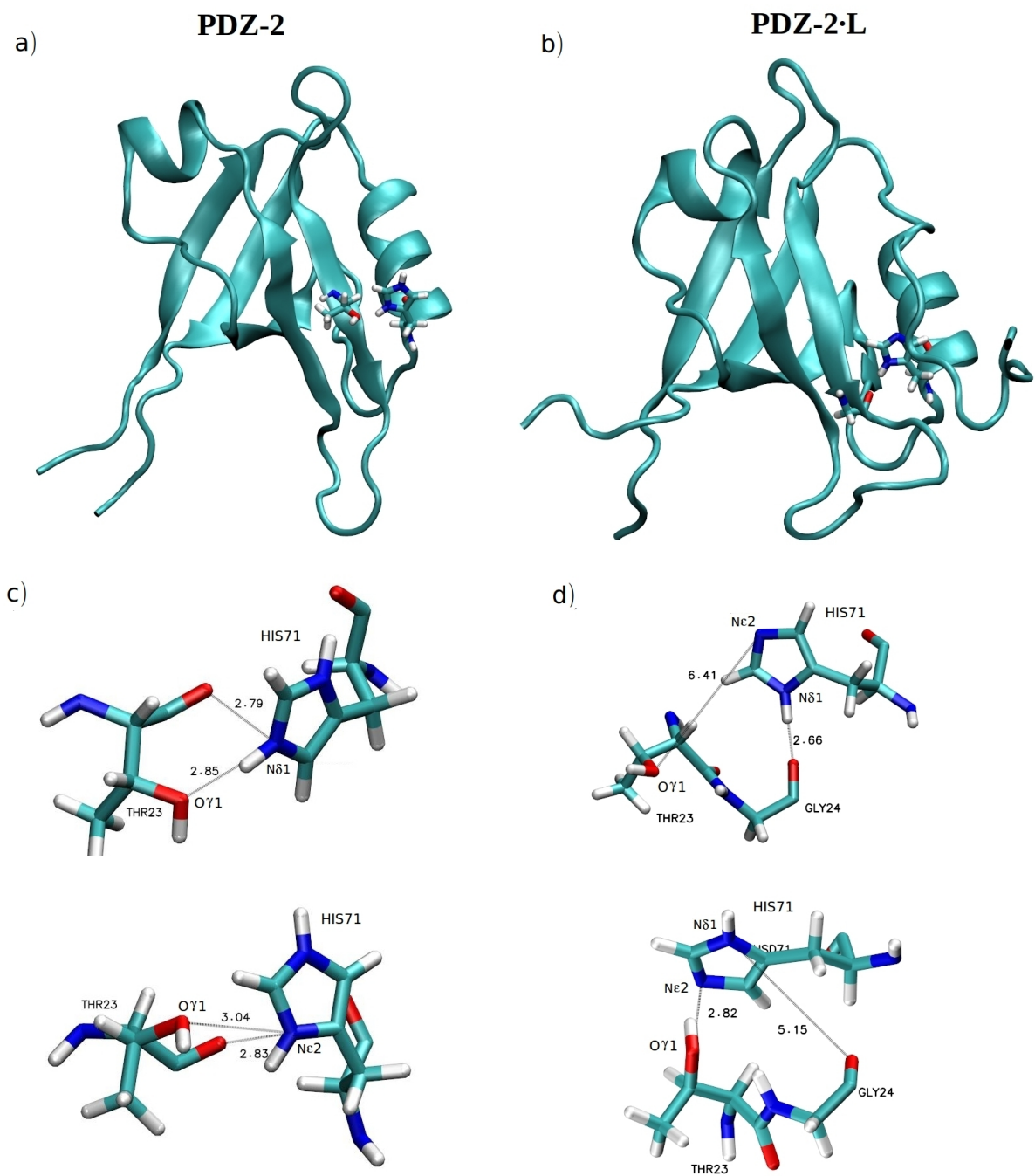

**Figure 3.** Snapshots of interaction for PDZ-2 (a) and PDZ-2·L (b), and details of the respective hydrogen bond interactions between His71-Thr23 (c) and His71-Thr23-Gly24 (d) that gave rise to $\beta_2$ and $\alpha_2$ coupling in each case.

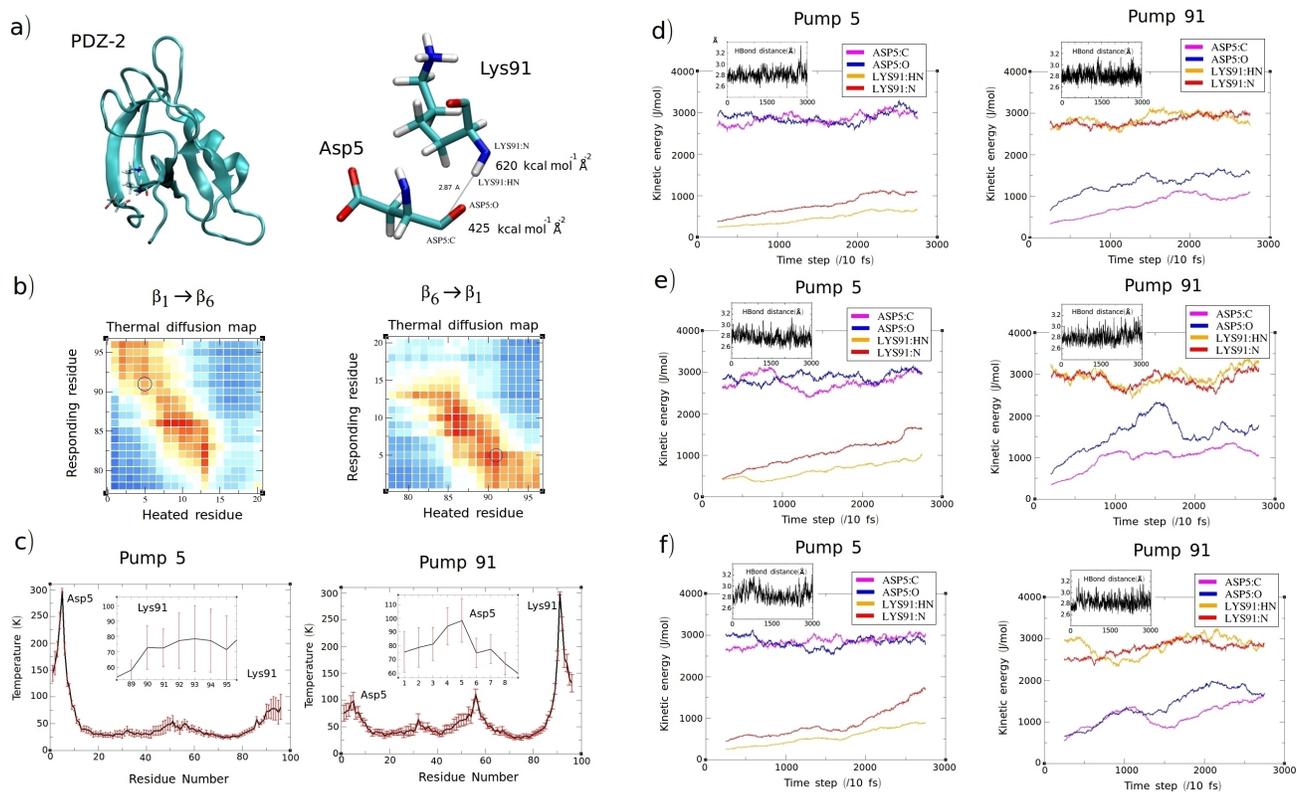

**Figure 4.** (a) Snapshot of Asp5-Lys91 interaction in PDZ-2 case. (b) Details of thermal maps for segments β1 - β6. (c) Pump-probe plots of Asp5 and Lys91 (n=20). (d,e,f) Kinetic analysis per atom, plotted using a running average of 250 time-steps. Insets represents hydrogen bond distance versus time-step for Lys91-N-H⋯O=C-Asp5 interaction (see text for details).

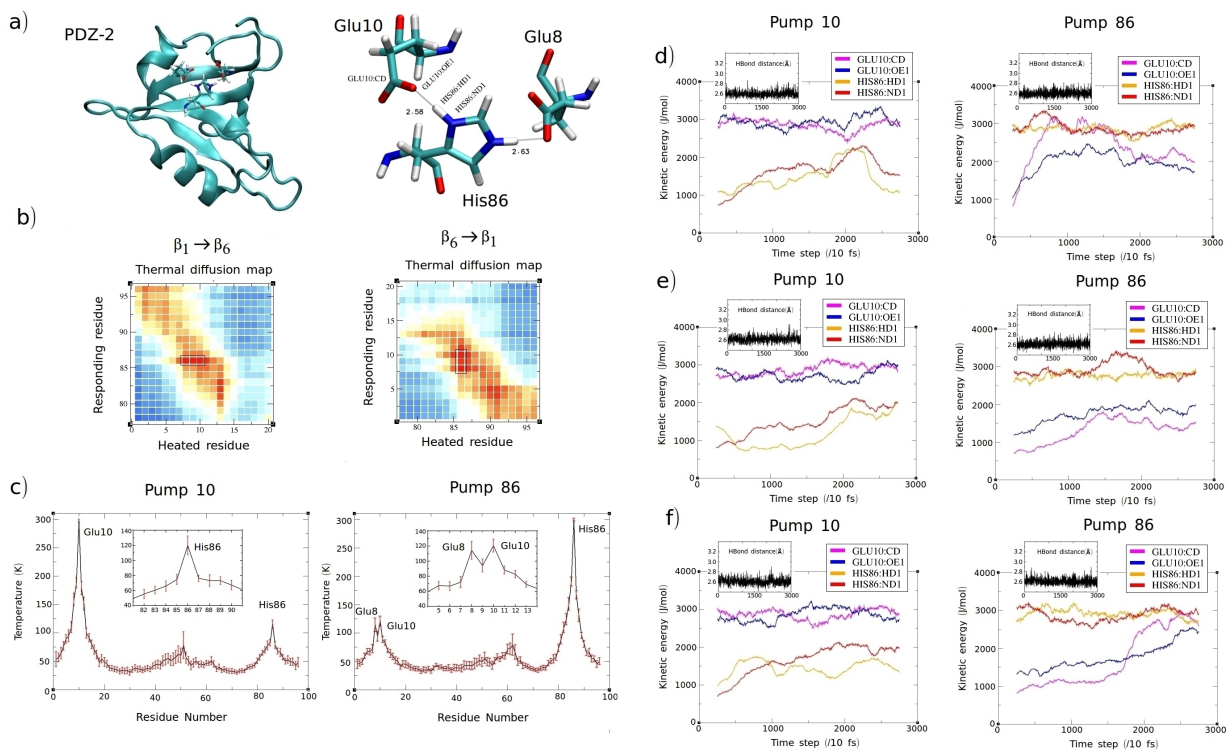

**Figure 5.** (a) Snapshot of Glu10-His86 interaction in PDZ-2 case. (b) Details of thermal maps for segments β1 - β6. (c) Pump-probe plots of Glu10 and His86 (n=20). (d,e,f) Kinetic analysis per atom, plotted using a running average of 250 time-steps. Insets represents hydrogen bond distance versus time-step for His86-N-H···O=C-Glu10 interaction (see text for details).

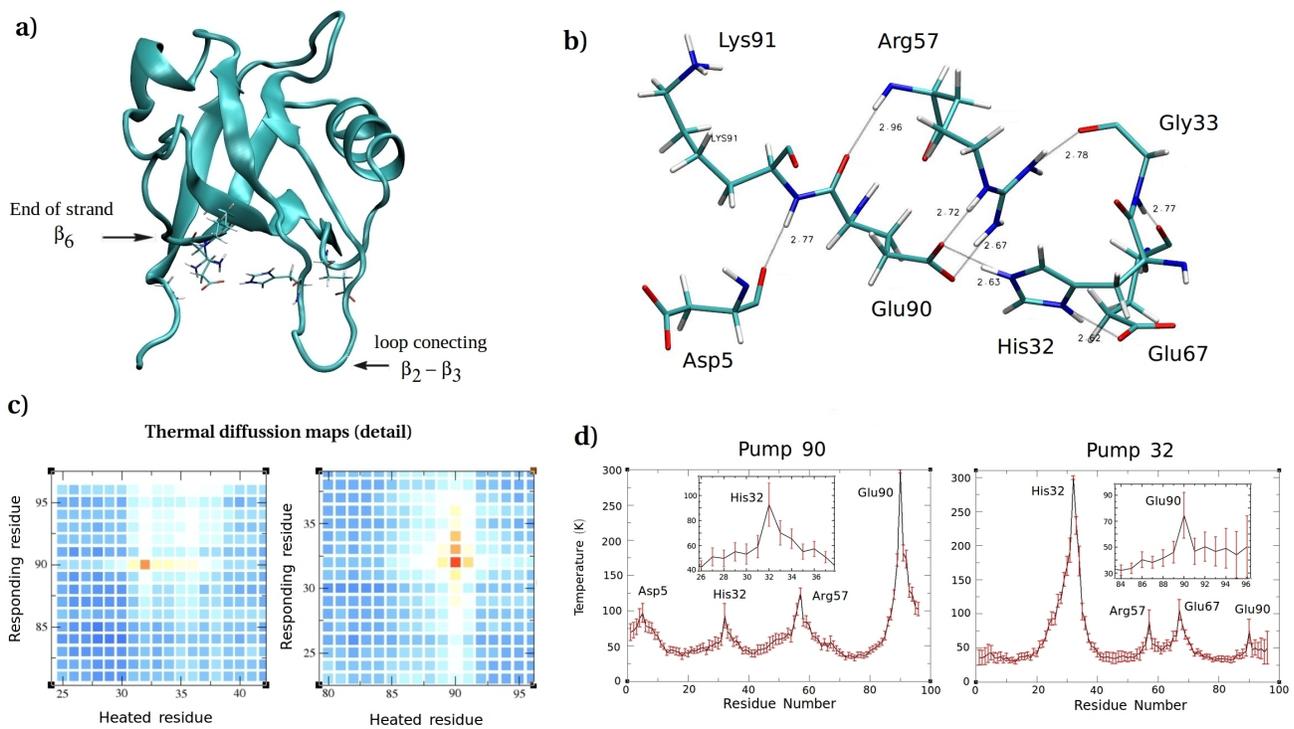

**Figure 6.** a) Structural location of residues Asp5, His32, Arg57, Glu67, and Glu90 in PDZ-2. b) Two details of relative residue orientations. c) Detail of thermal diffusion maps for the asymmetrical relationship between Glu90 and His32 (n=20). d) Averaged pump-probe of Glu90 and His32.

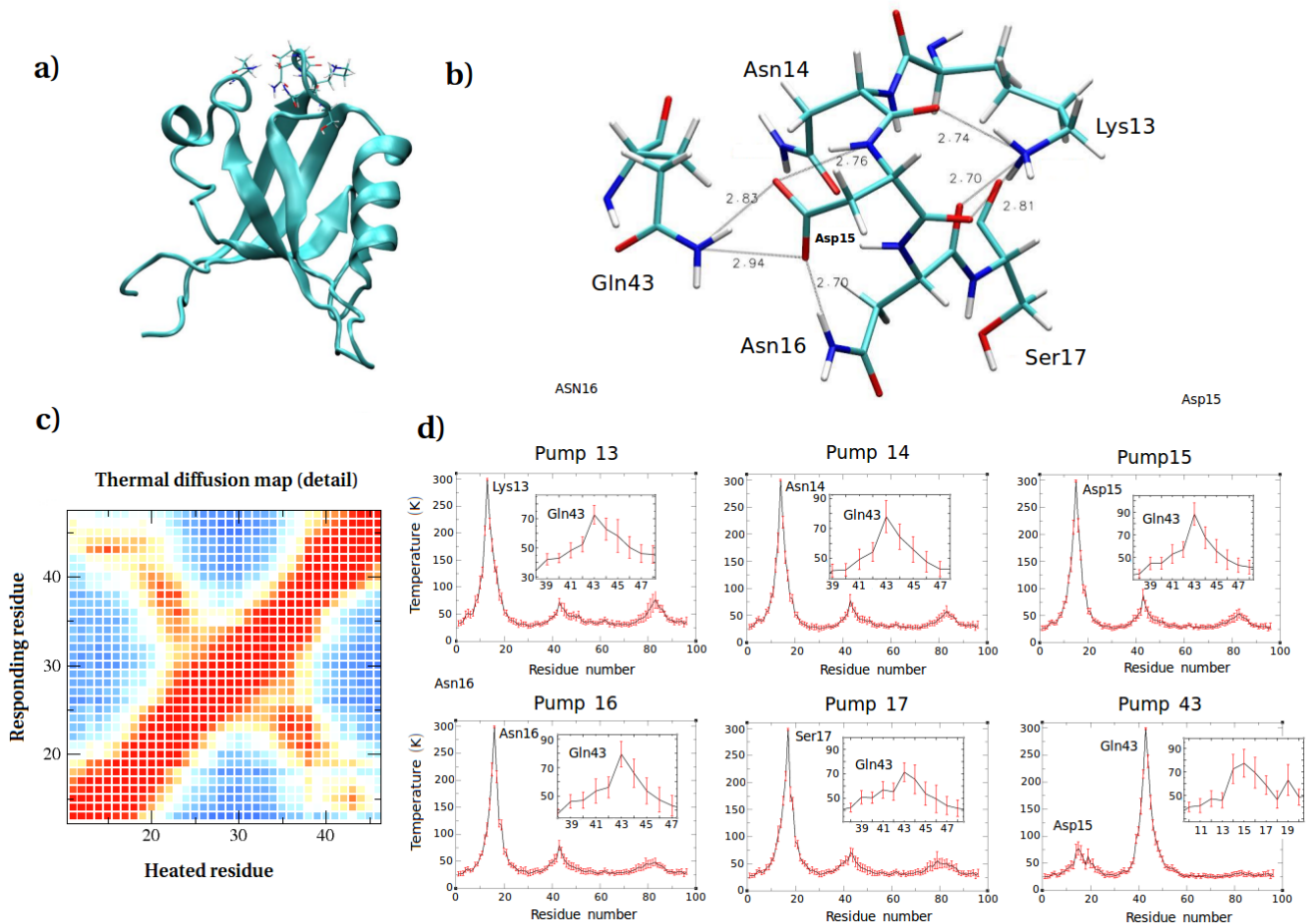

**Figure 7.** a) Structural location of residues Lys13, Asn14, Asp15, Asn16, Ser17, and Gln43 in PDZ-2·L. b) Detail of residue positions with hydrogen bond distances in Angstroms. c) Detail of the PDZ-2·L thermal map for the section involving Pathway I. d) Pump-probe relation for the involved residues (n=20).

**Graphical Abstract.**

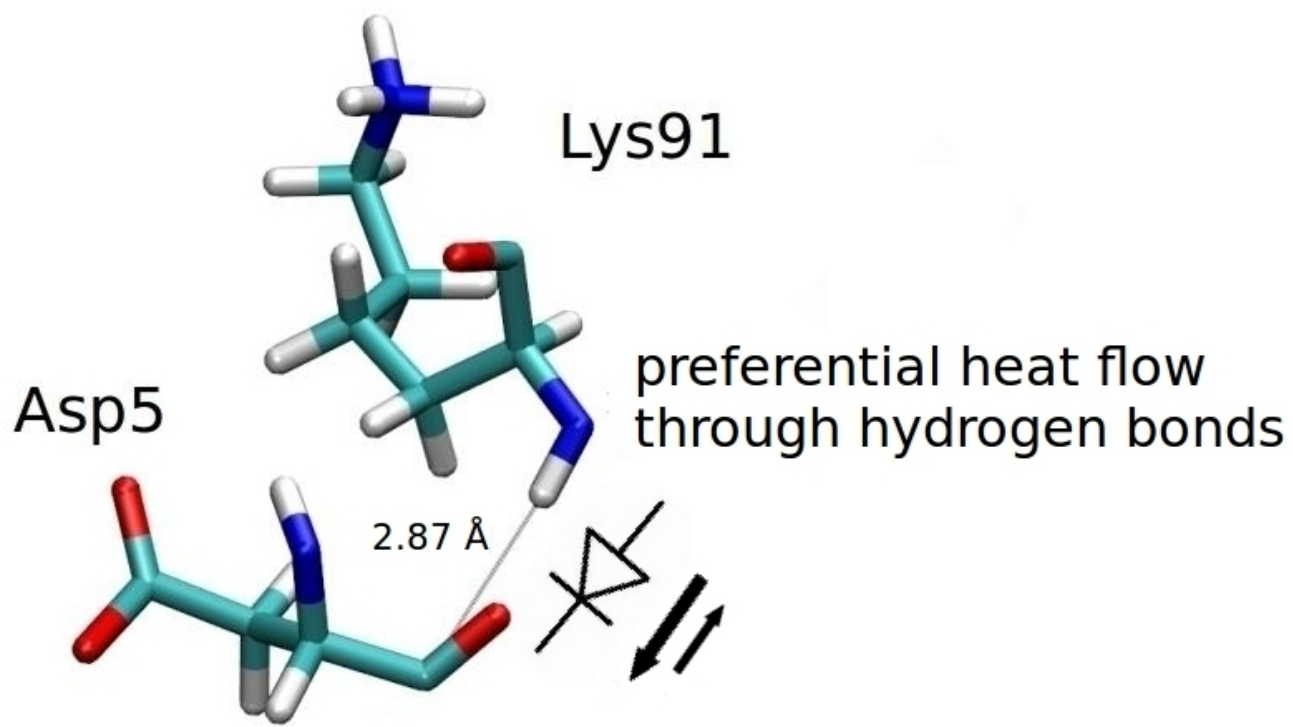